\documentclass[pbl,aps,twocolumn,superscriptaddress,shopacs,floatfix]{revtex4}

\usepackage{graphics,bm}
\usepackage{amssymb,ulem,amsmath}
\usepackage{epsfig}
\usepackage{epsf}
\usepackage{color}
\usepackage[colorlinks=true,urlcolor=blue,citecolor=blue, linkcolor = blue]{hyperref}
\newcommand{\bra}[1]{\langle #1|}
\newcommand{\ket}[1]{|#1\rangle}

\begin{document}

\title{Short-time Spin Dynamics in Strongly Correlated Few-fermion Systems}
\author{Sebastiano Peotta}
\affiliation{NEST, Scuola Normale Superiore and Istituto di Nanoscienze-CNR, I-56126 Pisa, Italy}
\author{Davide Rossini}
\affiliation{NEST, Scuola Normale Superiore and Istituto di Nanoscienze-CNR, I-56126 Pisa, Italy}
\author{Pietro Silvi}
\affiliation{International School for Advanced Studies (SISSA), Via Bonomea 265, 34136 Trieste, Italy}
\author{G. Vignale}
\affiliation{Department of Physics and Astronomy, University of Missouri, Columbia, Missouri 65211, USA}
\author{Rosario Fazio}
\affiliation{NEST, Scuola Normale Superiore and Istituto di Nanoscienze-CNR, I-56126 Pisa, Italy}
\author{Marco Polini}
\affiliation{NEST, Istituto di Nanoscienze-CNR and Scuola Normale Superiore, I-56126 Pisa, Italy}

\begin{abstract}
The non-equilibrium spin dynamics of a 
one-dimensional system of repulsively interacting fermions is studied by means of density-matrix renormalization-group simulations.  We focus on the short-time decay of the oscillation amplitudes of the centers of mass of spin-up and spin-down fermions. Due to many-body effects, the decay is found to evolve from quadratic to linear in time, and eventually back to quadratic as the strength of the interaction increases. The characteristic rate of the decay increases linearly with the strength of repulsion in the weak-coupling regime, while it is inversely proportional to it in the strong-coupling regime. Our predictions can be tested in experiments on tunable ultra-cold few-fermion systems in one-dimensional traps.
\end{abstract}
\maketitle

\section{Introduction}
In an electron liquid the motion of one of the two spin species, {\it e.g.} in the presence of a spin current, 
can drag along the other one because of electron-electron interactions. This is the spin Coulomb drag effect or simply the 
{\it Spin Drag} (SD)~\cite{scd_giovanni,scd_flensberg,polini_physics_2009}.  In electron transport  SD can be described 
by a frictional force proportional to the difference between the velocities of the two populations and is described by a 
damping term in the equation of motion for the time derivative of the spin-resolved center-of-mass momentum. 
SD has been observed~\cite{weber_nature_2005,yang_nature_2011} in two-dimensional electron gases
in semiconductor heterojunctions.

The concept of SD can be extended to other quantum fluids with distinguishable species that can exchange momentum due to mutual 
collisions. Ultracold atomic gases~\cite{reviewscoldatoms} are  clean systems 
in which SD can be observed in a truly {\it intrinsic} regime~\cite{polini_prl_2007,bruun_prl_2008,gao_prl_2008,duine_prl_2010,duine_prl_2009}. 
Further, the interaction strength between atoms can be tuned at will by employing Feshbach resonances~\cite{reviewscoldatoms}.

This work is motivated by a recent pioneering experiment~\cite{sommer_nature_2011} on SD in an equal mixture of two hyperfine states of $^6{\rm Li}$ atoms confined in a trap. 
The authors of Ref.~\onlinecite{sommer_nature_2011} measured independently the time-dependent position of the centers of mass of ``spin-up" and ``spin-down" particles starting from an initial condition in which the two types of particles are grouped in well-separated clouds.  The  experiment is performed in 
the ``unitarity limit" in which the strength of interactions is the largest possible.  At long times the separation of the centers of mass decays exponentially to zero.  By measuring the time constant of this exponential decay the SD coefficient is determined.

Besides providing information on SD in the strong coupling regime, Ref.~\onlinecite{sommer_nature_2011} provides a wealth of new data on the {\it short-time} behavior -- long before the SD regime is attained. There it is found that the two clouds perform several cycles of oscillation before settling at the bottom of the trap. If interactions are sufficiently strong, they reflect off each other several times before the inter-diffusion process begins. 
This short-time regime of spin dynamics, the short-time SD (STSD), constitutes the focus of the present work.
We tackle it non-perturbatively by the time-dependent 
density-matrix renormalization group (TDMRG) method~\cite{tdmrg} (see Sec.~\ref{appendix:TDMRG}). 
This method is essentially exact, its main limitation being the maximum system size that we can handle~\cite{recentstudies}.
Starting from an initial condition similar to that of Ref.~\onlinecite{sommer_nature_2011}, we find that the oscillation amplitudes of the centers of the spin clouds decay in time {\it quadratically} for weak interactions,  {\it linearly} for intermediate interactions, and again quadratically for very large interactions.
Below we argue that this intriguing reentrant behavior is a many-body effect.
Our predictions are amenable to experimental testing, since in a recent work Serwane {\it et al.}~\cite{serwane_science_2011} were able to trap few fermions in a 1D geometry and to tune their mutual interactions by means of a Feshbach resonance.

\section{The model}
 
We consider a two-component fermion system with repulsive short-range interactions in a 1D trap. 
The system is prepared in the ground state of  
two (spin-dependent) displaced harmonic potentials [Fig.~\ref{fig:one}a)]. At time $t=0^+$ these external potentials are {\it suddenly} 
turned off and the system evolves in presence of a single harmonic confinement, according to the Fermi-Hubbard (FH) Hamiltonian,
\begin{equation}\label{eq:hubbard}
	{\hat {\cal H}}=-J\sum_{i,\sigma}({\hat c}^{\dagger}_{i, \sigma}
	{\hat c}_{i+1,\sigma}+{\rm H}.{\rm c}.)+U\sum_i 
	{\hat n}_{i, \uparrow}{\hat n}_{i, \downarrow}+\sum_i W_i{\hat n}_i~.
\end{equation}
Here $J$ is the inter-site hopping parameter, ${\hat c}^{\dagger}_{i, \sigma}$ (${\hat c}_{i, \sigma}$) creates 
(destroys) a fermion in the $i$-th site ($i \in[1,L]$, $L$ being the total number of lattice sites), $\sigma=\uparrow,\downarrow$ 
is a label for a pseudospin-$1/2$ (hyperfine-state) degree of freedom, $U>0$ is the on-site repulsion, 
${\hat n}_{i, \sigma}={\hat c}^{\dagger}_{i, \sigma}{\hat c}_{i, \sigma}$ is the local spin-resolved  number operator, and 
${\hat n}_{i}= {\hat n}_{i, \uparrow} + {\hat n}_{i, \downarrow}$. The third term 
on the r.h.s. of Eq.~(\ref{eq:hubbard}) represents an external parabolic potential $W_i = V_2(i-L/2)^2$ of strength $V_2$, corresponding to a frequency $\omega = 2 \sqrt{V_2J}/\hbar$.

We follow the time-evolution of the spin-resolved densities $\langle {\hat n}_{i, \sigma}(t)\rangle$ 
on a time scale much smaller than the spin equilibration time~\cite{foot1} and calculate the spin-resolved centers of mass from $X_{{\rm CM}, \sigma}(t) \equiv L^{-1}\sum_{i =1}^L (i - L/2)\langle\Psi(t)|{\hat n}_{i, \sigma}|\Psi(t)\rangle$.

\section{Numerical results and discussion}
\label{sec:numerical_results}

\begin{figure}
\centering
\includegraphics[width=1.00\linewidth]{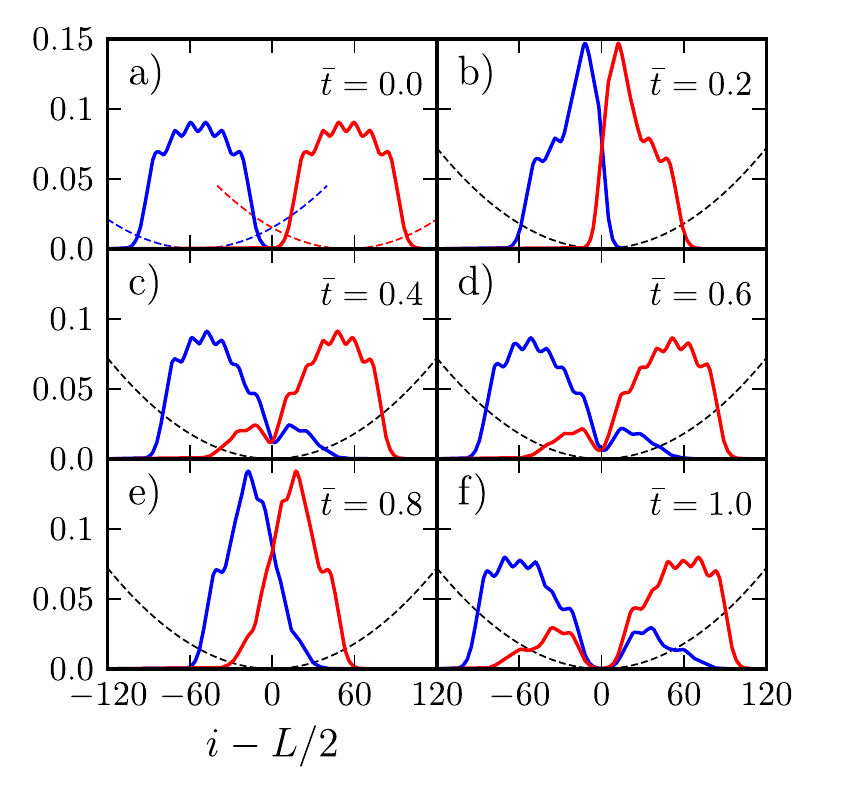}
\caption{(Color online) Time evolution of the site occupations $n_{i, \sigma}(t)$ for a system of twelve fermions at strong coupling ($U/J =5$). Panel a) Initial state: two clouds of atoms with opposite spin (blue and red curves) are spatially separated using two displaced harmonic potentials. Panels b)-f) Subsequent time evolution after the abrupt switch-off of these local potentials. The two clouds are forced to propagate against each other in presence of an overall harmonic confinement of frequency $\omega$. The parameter ${\bar t}$ denotes time in units of the period $T = 2\pi/\omega$ induced by the harmonic confinement. Dashed lines indicate the initial spin-dependent displaced harmonic traps [panel a)] and the overall harmonic trap for $t>0$ [panels b)-f)], and are plotted as guides to the eye. \label{fig:one}}
\end{figure}

 In Fig.~\ref{fig:one} we illustrate the time evolution of the occupation 
numbers $n_{i, \sigma}(t)$ for a system of $N = 6$ spin-up particles and $N=6$ spin-down particles, 
in a lattice with $L =240$ sites.  The harmonic potential has a strength $V_2/J = (1/160)^2$, corresponding to a harmonic oscillator length ${\bar a}_{\rm ho} = (J/V_2)^{1/4} \approx 12.65$, 
in units of the lattice constant, and to a frequency $\omega = J/(80 \hbar)$. 
These parameters have been used also for all other plots and their choice yields minimal lattice effects (see below)~\cite{foot2}. 
The data in Fig.~\ref{fig:one} correspond to $U/J = 5$. In panel a) 
we illustrate the initial state, with two non-overlapping clouds with opposite 
spins. Panels b)-f) show the time evolution of this initial state. We highlight two features: i) in panels b) and e) high-density 
regions form near the center of the trap due to strong repulsive interactions~\cite{sommer_nature_2011}; 
ii) in panels c), d), and f) we see how the spin-up cloud (blue curve) drags along a substantial fraction of down-spin atoms (red curve).

\begin{figure}
\centering
\includegraphics[width=1.00\linewidth]{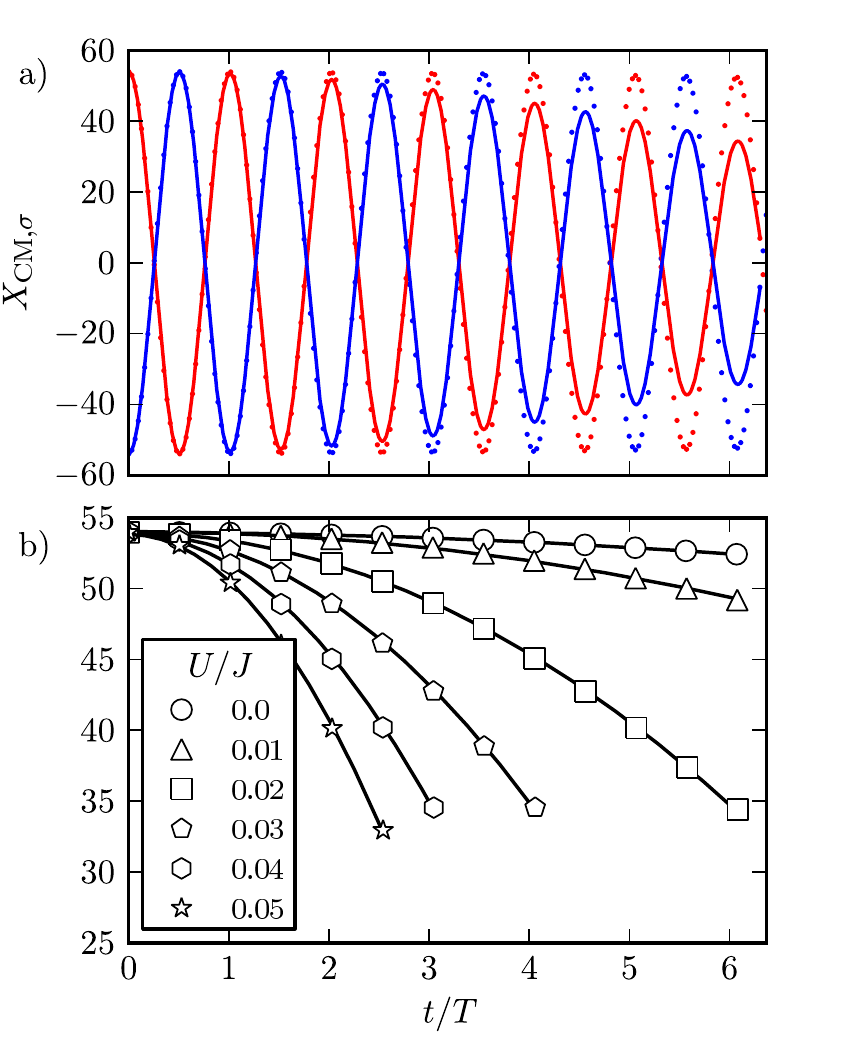}
\caption{(Color online) Panel a) Time evolution of the spin-resolved center-of-mass $X_{{\rm CM}, \sigma}(t)$ of a system of twelve fermions in a harmonic potential. Solid lines refer to $U/J =0.02$ while dotted lines to $U/J=0$. Panel b) Positions of the maxima of the amplitude of the center-of-mass oscillations as functions of time $t$ in units of $T = 2\pi/\omega$. Different symbols correspond to different interaction strengths. The tiny decay in the non-interacting case is due to lattice effects. Solid lines are parabolic fits, $\left. X_{{\rm CM}, \sigma}(t)\right|_{\rm peak} = X_0~[1 - (t/\tau_{\rm STSD})^2]$, where $X_0$ is the same for all values of $U/J$.
\label{fig:two}}
\end{figure}

In Fig.~\ref{fig:two}a) we show the time evolution of spin-resolved center-of-mass, $X_{{\rm CM}, \sigma}(t)$, {\it in the weak coupling regime}, $U/J \leq 0.05$. 
In absence of the lattice, the center-of-mass of each atomic cloud is decoupled from ``internal" degrees of freedom and should oscillate at the trap frequency, $\omega$, without decaying. 
This is confirmed by the data corresponding to $U/J =0$ in Fig.~\ref{fig:two}a) (dotted lines).  No visible damping effects appear 
within the time-scale of the plot, since we have minimized lattice effects~\cite{rey_pra_2005}.

When $U/J$ is finite the two clouds still go through each other, but their motion is damped.  Fig.~\ref{fig:two}b) reports the maxima of the blue and red curves as 
a function of time, for several different values of $U/J \leq 0.05$. The amplitude of the oscillations in Fig.~\ref{fig:two}a) decays quadratically in time. 
This is because, in  this regime, the center of mass of each cloud is a harmonic oscillator weakly coupled to internal degrees of freedom.  The relevant excitation spectrum, ${\cal S}(\Omega)$, is sharply peaked about $\Omega \sim \omega$. The position of the peak determines the frequency of the oscillations and the second moment of the spectrum determines the quadratic decay of their amplitude. 
The quadratic decay can be also verified analytically by means of time-dependent perturbation theory -- see Sec.~\ref{appendix:PT}. 
\begin{figure}
\centering
\includegraphics[width=1.00\linewidth]{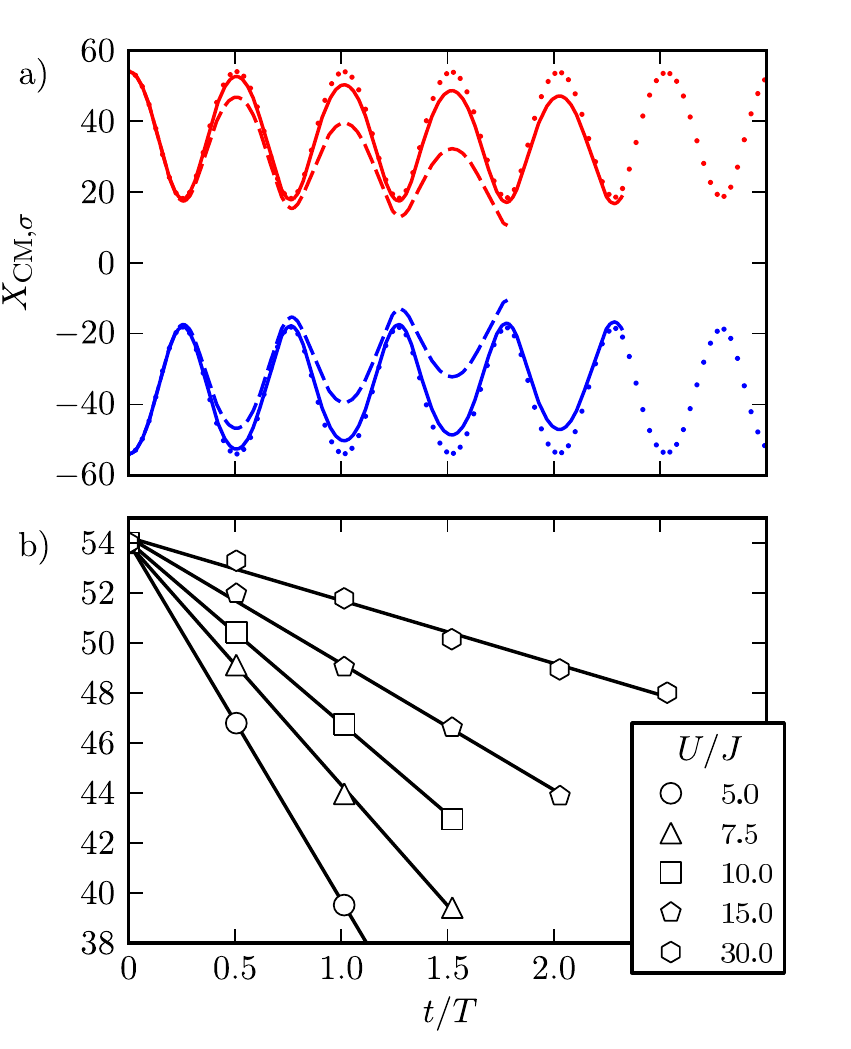}
\caption{(Color online) Panel a) Same as in Fig.~\ref{fig:two}a) but at strong coupling. Solid (dashed) lines refer to $U/J =20$ ($U/J=5$) while dotted lines to $U/J =\infty$. In the latter case the oscillations have twice shorter periodicity than those of non-interacting fermions. Note that they do not display an appreciable decay on the time scale of the plot. Panel b) Same as in Fig.~\ref{fig:two}b) but at strong coupling. 
Solid lines are fits obtained by using Eq.~(\ref{eq:splitfit}) in Sec.~\ref{appendix:tstar}.\label{fig:three}}
\end{figure}

\begin{figure*}
\centering
\includegraphics[width=1.00\linewidth]{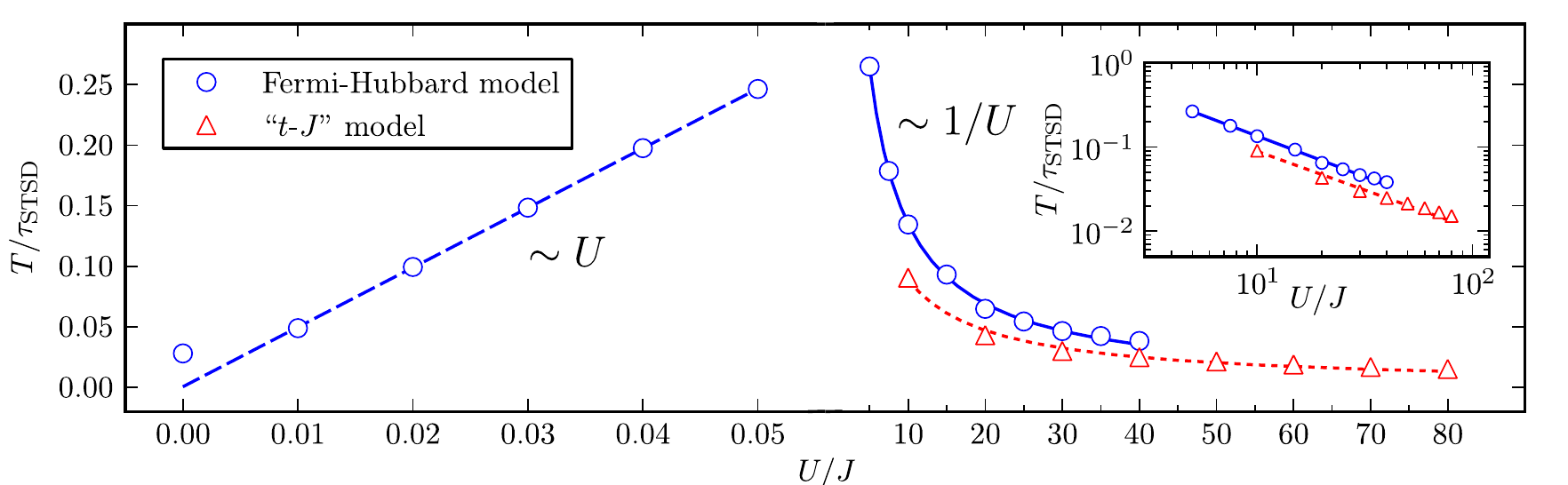}
\caption{(Color online) 
Empty circles represent the inverse of the short-time spin-drag time constant, $\tau^{-1}_{\rm STSD}$ (in units of $1/T$), as a function of the coupling $U/J$.  These data have been extracted from the spin dynamics of the model (\ref{eq:hubbard}) (quadratic fit at weak coupling and ``split-fit'' procedure at strong coupling, see Sec.~\ref{appendix:tstar}). In a wide range of coupling constants, $0.1 \lesssim U/J \lesssim 1$, fittings like the ones  
in Figs.~\ref{fig:two}b)-\ref{fig:three}b) do not work. We see that $\tau^{-1}_{\rm STSD}$ vanishes linearly in the weak-coupling regime (long-dashed line) and behaves approximately like $1/U$ at strong coupling (solid line). 
The solid line is a power-law fit, {\it i.e.} $1/\tau_{\rm STSD} = A~(U/J)^{-\alpha}$, with $A\approx 1.26/T$ and $\alpha \approx 0.97$. 
The empty triangles label $\tau^{-1}_{\rm STSD}$ as extracted from the spin dynamics of the effective model (\ref{eq:tJ}). The short-dashed line is a power-law fit of the form $1/\tau_{\rm STSD} = B~(U/J)^{-\beta}$, with $B \approx 0.7/T $ and $\beta \approx 0.91$. 
While the two exponents $\alpha$ and $\beta$ are very similar, the proportionality constants $A$ and $B$ are slightly different. 
This discrepancy may be due to the neglect of three-site terms~\cite{mapping} in Eq.~(\ref{eq:tJ}).
In the inset we show the same strong-coupling results in a log-log scale. \label{fig:four}
}
\end{figure*}

We now discuss the strongly correlated regime,  $U/J \gg 1$. 
The main results are summarized in Fig.~\ref{fig:three}. Dotted lines in Fig.~\ref{fig:three}a) represent the exact time evolution 
of the spin-resolved center-of-mass for $U/J =\infty$. In this limit, the Hamiltonian in Eq.~(\ref{eq:hubbard}) maps onto~\cite{giamarchi_book}
${\hat {\cal H}}_{\infty} = -J{\hat {\cal P}} \sum_{i,\sigma}({\hat c}^{\dagger}_{i, \sigma}
	{\hat c}_{i+1, \sigma}+{\rm H}.{\rm c}.){\hat {\cal P}}~,
$
where ${\hat {\cal P}}$ is a Gutzwiller projector (that avoids double occupation of a lattice site). Dotted lines  
have been obtained by applying TDMRG to ${\hat {\cal H}}_{\infty}$. Notice that the centers of mass of the two clouds behave 
practically like two classical particles that bounce off each other quasi-elastically oscillating at twice the trap frequency. 
The frequency doubling with respect to Fig.~\ref{fig:two}a) is understandable as follows.  Due to strong repulsion, fermions
of opposite spin are confined to one half of the trap: effectively, only the antisymmetric levels of the harmonic oscillator, whose energy separation is $2\omega$, are involved in the time evolution. A rather complicated dynamical pattern, however, is present in the time evolution of the spin-resolved site occupations $n_{i, \sigma}(t)$ (see Fig.~\ref{fig:one}). 

In Fig.~\ref{fig:three}b) we plot the amplitude of oscillations vs time  for $5 \leq U/J \leq 30$.  Remarkably they decay {\it linearly}. 
As mentioned in the Introduction, the quadratic-to-linear crossover is a many-particle effect. One can indeed solve analytically the evolution dynamics for an interacting system of {\it two} particles with antiparallel spin in a harmonic potential. In that case, the time evolution of $X_{{\rm CM}, \sigma}$ follows the quadratic behavior seen in Fig.~\ref{fig:two}b), even for strong interactions (see Sect.~\ref{appendix:twobody}). 
With many particles, as the strength of the interaction increases, the centers of mass of the clouds become increasingly coupled to internal degrees of freedom. If $N$ is sufficiently large, ${\cal S}(\Omega)$ becomes featureless, with a bandwidth of the order of $\omega$.  In this regime one has the situation of a single degree of freedom (center of mass) irreversibly transferring energy into a ``bath" of microscopic degrees of freedom: accordingly, the amplitude of the oscillations decays linearly in time as expected of an ordinary damped oscillator. 

These observations imply a non-trivial crossover in the short-time dynamics of $\left. X_{{\rm CM}, \sigma}(t)\right|_{\rm peak}$ 
as a function of the number of particles. In particular, as illustrated in Figs.~\ref{fig:five-SOM}-\ref{fig:eight-SOM} in Sec.~\ref{appendix:tstar}, we note the existence of a time scale $t^\star$, depending on $N$ and $U/J$, below which the decay of the oscillation amplitudes is {\it quadratic}. The value of $t^\star$ decreases with increasing $N$ and increases with increasing $U/J$. 
More quantitatively, we have investigated such crossover by fitting numerical data at strong coupling with the ``split-fit'' formula in Eq.~(\ref{eq:splitfit}) of Sec.~\ref{appendix:tstar}, which contains $\tau_{\rm STSD}$ and $t^\star$ as fitting parameters. This equation encodes a quadratic decay for $t \leq t^\star$, followed by a linear behavior for $t > t^\star$. From our analysis we conclude that the value of $t^\star$ for $N=6$ and $5 \lesssim U/J \lesssim 30$ is much smaller than the period of oscillations. This explains why no quadratic behavior is  seen in Fig.~\ref{fig:three}b). 
From the numerical data at strong coupling, we conclude that a linear decay in time of $\left. X_{{\rm CM}, \sigma}(t)\right|_{\rm peak}$ occurs when the overlap between the two colliding clouds is substantial, while a quadratic decay takes place initially (for $t < t^\star$) when minor overlap occurs in the tails of the clouds. 

Our main results for the time scale $\tau_{\rm STSD}$ associated with STSD are summarized in Fig.~\ref{fig:four}. Here 
we report the values of $\tau^{-1}_{\rm STSD}$ used to produce the fits in Figs.~\ref{fig:two}b)-\ref{fig:three}b). We clearly see that $\tau^{-1}_{\rm STSD}$ vanishes linearly in the weak-coupling $U/J \to 0$ limit. This  
has to be contrasted with the SD relaxation rate in a system with many degrees of freedom at equilibrium: the latter is 
quadratic in the coupling constant governing the strength of inter-particle interactions (see {\it e.g.} Ref.~\onlinecite{polini_prl_2007}). 
In the strong-coupling limit $\tau^{-1}_{\rm STSD}$ behaves approximately 
like $1/U$. No analytical results are available in this regime, even in a system with many degrees of freedom at equilibrium. 
We emphasize that, for all the data at $U/J \gg 1$ in Fig.~\ref{fig:four}, $t^\star$ is within the observation time of our simulations.

To check the robustness of our conclusions at strong coupling, we study an effective ``$t$-$J$" model~\cite{giamarchi_book,essler_book} which approximates (\ref{eq:hubbard}) for 
$U/J \gg 1$:
\begin{equation}\label{eq:tJ}
{\hat {\cal H}}' =
{\hat {\cal H}}_{\infty} +
 \frac{4J^2}{U}\sum_{i} \left({\hat {\bm S}}_i \cdot {\hat {\bm S}}_{i+1} - \frac{{\hat n}_{i} {\hat n}_{i+1}}{4}\right)
 +\sum_i W_i{\hat n}_i~,
\end{equation}
where ${\hat {\bm S}}_i = \sum_{\alpha} {\hat c}^\dagger_{i,\alpha} ({\bm \sigma}_{\alpha\beta}/2) {\hat c}_{i,\beta}$ is the 
spin operator (${\bm \sigma}$ being a three-dimensional vector of Pauli matrices)~\cite{mapping}. 
When $U/J \gg 1$ such model is much easier to simulate than the original FH model.  Employing Eq.~(\ref{eq:tJ}) we have discovered that, for a fixed value of $N$, the amplitude of the oscillations decays quadratically in time when $U/J$ is sufficiently large. This is shown in Fig.~\ref{fig:four-SOM} in Sec.~\ref{appendix:tstar}. 
 In other words, as mentioned in the Introduction, the quadratic dependence on time of the decay of the oscillation amplitudes displays a reentrant behavior pertaining to the many-particle problem. In Fig.~\ref{fig:four} we report the results for the inverse STSD time constant of the model (\ref{eq:tJ}) (empty triangles), which agree qualitatively with those based on the full FH model.

In summary, we studied short-time spin-density oscillations in a strongly-interacting 1D few-fermion system. 
We discovered that the decay in the oscillation amplitudes goes from quadratic to linear back to quadratic in time as the interaction strength increases from zero to infinity. 
The inverses of the properly-defined time constants depend on the strength of inter-particle interactions in a way that was unpredictable 
on the basis of our knowledge of the same phenomenon in many-particle systems near equilibrium. 
Our predictions can be tested by studying the damping of spin-dipole oscillations in few-fermion systems~\cite{serwane_science_2011}.

\acknowledgements We gratefully acknowledge financial support by the EU FP7 Programme under Grant Agreement 
No. 248629-SOLID, No. 234970-NANOCTM, and No. 215368-SEMISPINNET and by the NSF under Grant No. DMR-1104788.

\appendix
\begin{figure*}[t]
\centering
\begin{minipage}[t]{.48\textwidth}
\includegraphics[scale=1]{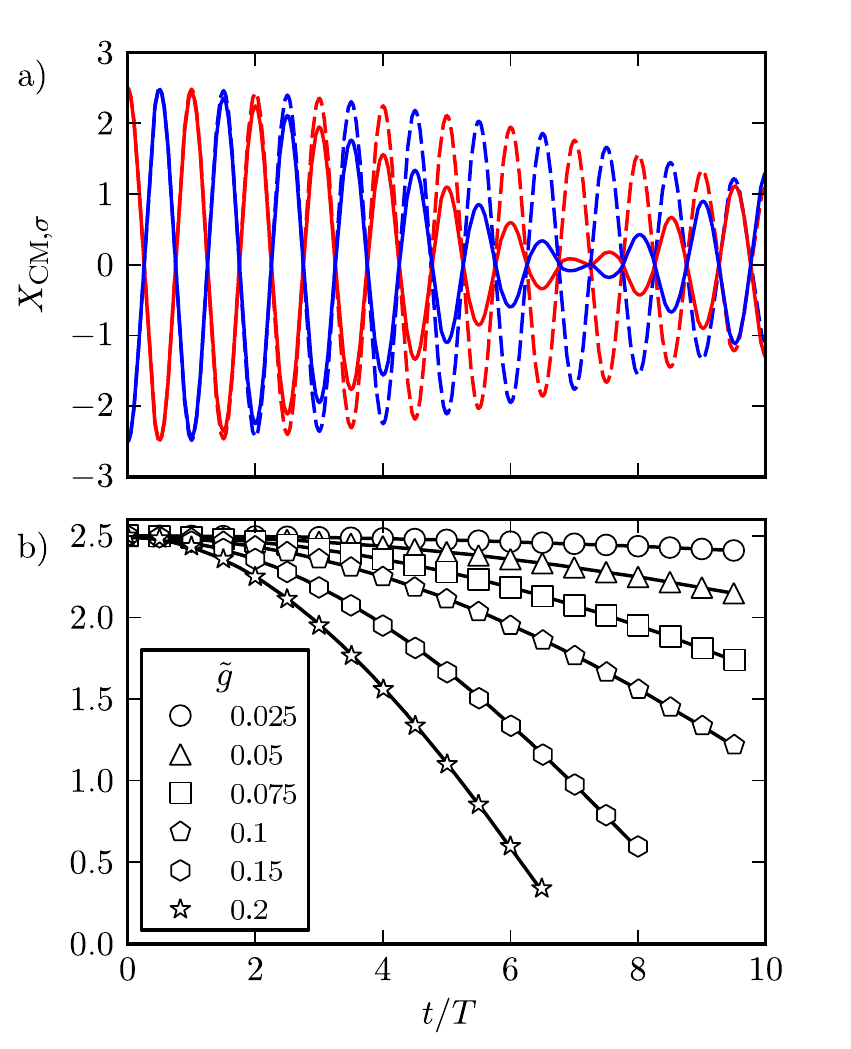}
\caption{(Color online) Panel a) Time evolution of the spin-resolved center-of-mass $X_{{\rm CM}, \sigma}(t)$ of a system of two fermions in a harmonic potential. Solid lines correspond to $\tilde{g}=0.2$, while dashed ones to ${\tilde g}=0.1$. Panel b) Positions of the maxima of the amplitude of the center-of-mass oscillations as functions of time $t$ in units of $T = 2\pi/\omega$. Different symbols label data corresponding to different values of the coupling constant ${\tilde g}$. The solid lines are fits of the form 
$\left. X_{{\rm CM}, \sigma}(t)\right|_{\rm peak} = X_0~\cos{(\sqrt{2}~t/\tau_{\rm STSD})}$, where $X_0$ is the same for all values of ${\tilde g}$.\label{fig:one-SOM}}
\end{minipage}
\hspace{5mm}
\begin{minipage}[t]{.48\textwidth}
\includegraphics[scale=1]{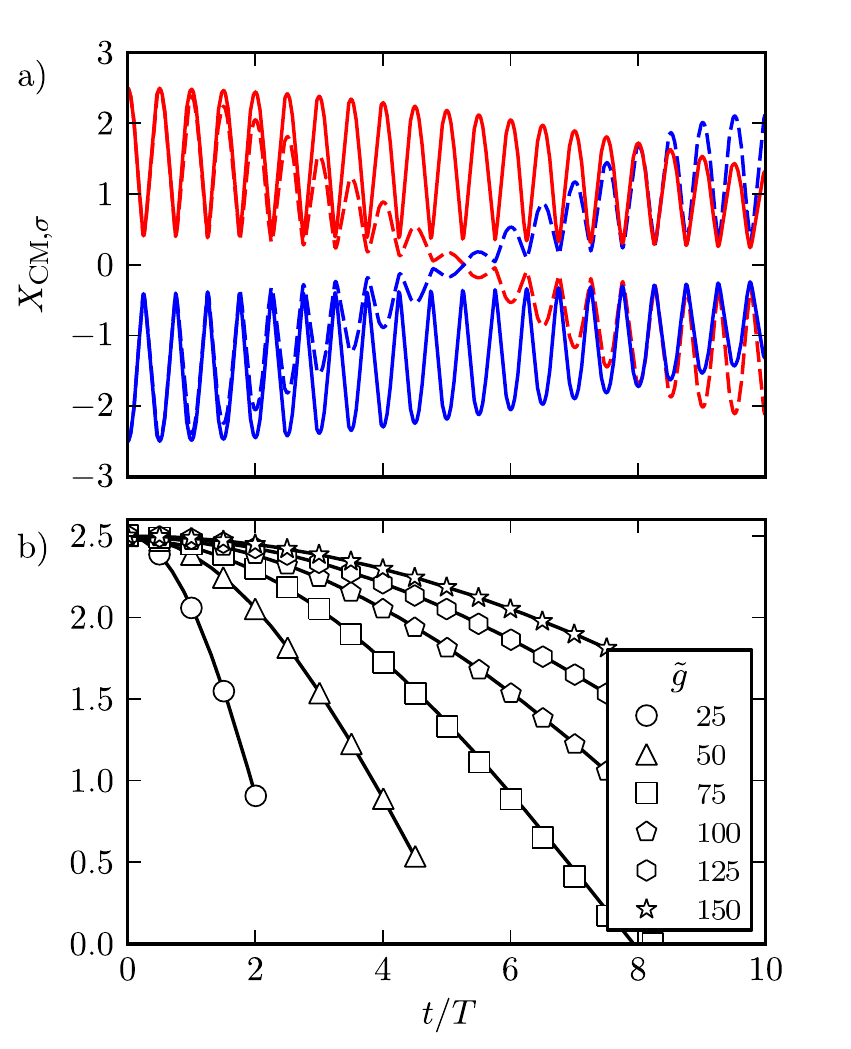}
\caption{(Color online) Same as Fig.~\ref{fig:one-SOM} but in the strong-coupling regime. Solid lines in panel a) correspond to ${\tilde g} = 150$, while dashed ones correspond to ${\tilde g}=50$. We stress that the fitting function used in panel b) is the same as in panel b) of Fig.~\ref{fig:one-SOM}, {\it i.e.} $\left. X_{{\rm CM}, \sigma}(t)\right|_{\rm peak} = X_0~\cos{(\sqrt{2}~t/\tau_{\rm STSD})}$.\label{fig:two-SOM}}
\end{minipage}
\end{figure*}
\begin{figure}[t]
\begin{center}
\includegraphics[scale = 1]{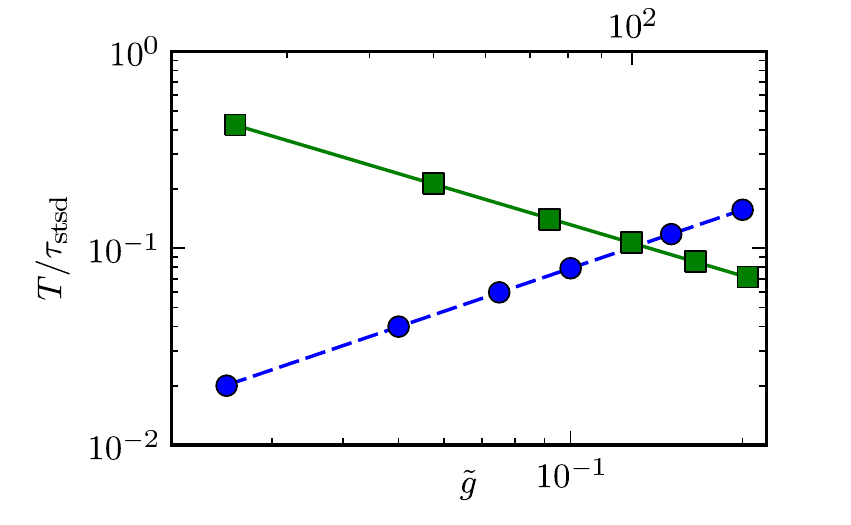}
\caption{(Color online) Inverse of the STSD time scale $\tau_{\rm STSD}$ as a function of the coupling constant ${\tilde g}$. All axes are in logarithmic scale. Data labeled by circles refer to the weak-coupling regime (lower horizontal axis), while the ones labeled by squares refer to the strong-coupling regime (upper horizontal axis). The solid line is a power-law fit, $1/\tau_{\rm STSD} = A~{\tilde g}^{-\alpha}$ with $A\approx 10.4/T$ and $\alpha \approx 0.99$. The dashed line is a power-law fit, $1/\tau_{\rm STSD} = C~\tilde{g}^{\gamma}$ with $C \approx 0.77/T$ and $\gamma \approx 0.99$.\label{fig:three-SOM}}
\end{center}
\end{figure}
\section{Technical details on the numerical method\label{appendix:TDMRG}}

In this Appendix we give some technical details on the numerical method we have used. 
Our simulations are based on the TDMRG method, which is known~\cite{tdmrg} to be a powerful technique for the
simulation of 1D systems.

In this work we have used a matrix product state (MPS) representation 
of the wave function, enforcing separate conservation of the number of spin-up and spin-down particles. The ground state at $t = 0$ is found using the procedure described in Ref.~\onlinecite{white_correction} ({\it modulo} small variations).
Inversion symmetry with respect to the trap center, {\it i.e.} $n_{i, \sigma}(t) = n_{L/2-i, {\bar \sigma}}(t)$,  
is not enforced {\it a~priori} but is present in the converged results 
(we use this feature as one of the benchmarks of the simulations).  
The truncation step is treated in a fully ``dynamical" way, {\it i.e.} we do not fix a maximum (or a minimum) number 
of states $m$ for each bond link. On the other hand, we choose to discard states with a ``small" statistical weight, summing up to a maximum allowed error $\epsilon$. 
This represents the crucial parameter that controls the precision and the duration of our simulations. Typically, we use $\epsilon \sim 10^{-8}-10^{-10}$. We have checked that these values are 
sufficiently small by employing two different procedures: i) we have compared our numerical results with the exact solution that is
available in the non-interacting case $U=0$ and ii) we have checked the accuracy
in the interacting case $U>0$ by analyzing the convergence of the results with decreasing $\epsilon$.
With this dynamical-truncation procedure the maximum number of states used in the simulations is $m \sim 1 \times 10^3 - 5 \times 10^3$, which is reached at the trap center where the time-evolution of $n_{i,\sigma}$ is rather complex (as we have seen in Fig.~\ref{fig:one}). 

The time-evolution operator is treated within a Suzuki-Trotter expansion. 
Due to the presence of nearest-neighbor-only interactions, one can separate
couplings on odd bonds from couplings on even bonds, thus writing the global Hamiltonian
as the sum of two non-commuting terms, ${\hat {\cal H}} = {\hat {\cal H}}_{\rm even} + {\hat {\cal H}}_{\rm odd}$. 
Each of the two contributions is the sum of commuting two-site terms.
To the $n$-th order the time-evolution operator reads
\begin{equation}
e^{-i \hat{\mathcal{H}} \Delta t} = \prod_{j=1}^k  e^{-i \hat{\mathcal{H}}_{\rm even} c_j \Delta t} e^{-i \hat{\mathcal{H}}_{\rm odd} d_j \Delta t} 
+ O(\Delta t^{n+1})~,
\end{equation}
where the number $2k$ of exponentials to be multiplied as well as the coefficients $c_j$ and $d_j$ depend
on the order of the expansion~\cite{Yoshida_1990}.
In our simulations we employ a sixth-order Suzuki-Trotter expansion (which interestingly enough was found to perform faster than the second-order one).

\section{Perturbation theory in the weak-coupling regime\label{appendix:PT}}

In the weak-coupling limit it is possible to study the impact of a contact repulsive interaction 
on the spin-resolved center-of-mass dynamics by means of time-dependent perturbation theory. 
For the sake of simplicity, we consider a continuum model, which is more amenable to an analytic treatment ($\hbar =1$): 
\begin{eqnarray}\label{ham}
{\hat {\cal H}} &= &\int dx~\bigg[\sum_\sigma\hat\psi_\sigma^{\dagger}(x)\left(-\frac{1}{2m}\frac{d^2}{dx^2}+\frac{1}{2}m\omega^2x^2\right)\hat\psi_\sigma(x) \nonumber \\
&+& g{\hat\rho}_\uparrow(x) {\hat \rho}_{\downarrow}(x)\bigg]~.
\end{eqnarray} 
The operators ${\hat \psi}_\sigma(x)$ with $\sigma = \uparrow,\downarrow$ are anticommuting field operators 
for spin-up and spin-down particles, while $\hat\rho_{\sigma}(x) = \hat\psi_\sigma^{\dagger}(x) \hat\psi_\sigma(x)$ 
are spin-resolved density operators. 

The state of the system before the quench ($t \leq 0$) corresponds 
to $N$ spin-up and $N$ spin-down particles in the ground states of two separate spin-resolved harmonic confinements. 
The distance $2d$ between the harmonic traps is supposed to be large enough 
(a few harmonic oscillator lengths) so that the initial state $|\psi(t=0)\rangle$ is (to a very good approximation)
\begin{equation}
\ket{\psi(0)} = e^{-i \hat{P}_{\uparrow}d }e^{i \hat{P}_{\downarrow}d }\ket{N\uparrow} \otimes \ket{N\downarrow}~,
\end{equation}
where $\ket{N\sigma}$ is the non-interacting ground state of $N$ particles with spin $\sigma$ in a harmonic potential and 
the translation operator ${\hat P}_\sigma$ is
\begin{equation}
\hat{P}_\sigma = -i \int dx~\hat{\psi}_\sigma^{\dagger}(x)\frac{d }{dx}\hat{\psi}_\sigma(x)~.
\end{equation}
The time evolution of the system for $t > 0$ is dictated by the Hamiltonian \eqref{ham}.

We now expand the two exponentials in the (small) parameter $g t$, obtaining, up to first order, the following result
\begin{eqnarray}\label{eq:expansionX}
X_{{\rm CM}, \sigma}(t)  &=& \pm d\cos(\omega t) + g\int_0^{t} dt' \int dx' \int dx~\frac{x}{N}~\nonumber\\
&\times& D_{\sigma}(x,x', t - t') n_{\bar\sigma}(x' \pm 2d\cos(\omega t')) \nonumber\\
&+& O(g^2t^2)~,
\end{eqnarray}
where the plus (minus) sign refers to $\sigma = \uparrow$ ($\sigma = \downarrow$), ${\bar \sigma}$ denotes the spin component opposite to $\sigma$, and $n_\sigma (x) = \bra{N\sigma} \hat{\rho}_\sigma(x) \ket{N\sigma}$ is the spin-resolved ground-state density profile.
The quantity $D_\sigma$ in the second term in the r.h.s. of Eq.~(\ref{eq:expansionX}) is the density-density response function:
\begin{eqnarray}
D_{\sigma}(x, x', t - t') &=& - i \Theta(t - t') \nonumber\\
&\times &\bra{N\sigma}\left[\hat{\rho}_\sigma(x,t), \hat{\rho}_\sigma(x', t')\right]\ket{N\sigma}~,\nonumber\\
\end{eqnarray}
$\Theta(t)$ being the Heaviside step function.

It can be easily shown that the density-density response function for the harmonic oscillator has the following properties
\begin{equation}
\left\{
\begin{array}{l}
D_{\sigma}(x, x', t - t')  =  D_{\sigma}(-x,-x', t - t')\vspace{0.1 cm}\\
D_{\sigma}(x, x', t - t') =  D_{\sigma}(x, x', t - t' +T)
\end{array}
\right.
~,
\end{equation}
where $T=2\pi/\omega$. Moreover, since the first argument of $D_\sigma(x,x', t-t')$ is integrated after being multiplied by an odd function [$x$ in the first line of Eq.~(\ref{eq:expansionX})], 
only the antisymmetric part of the response function is needed:
\begin{eqnarray}
D^{\rm (A)}_{\sigma}(x, x', t - t') &=& - D^{\rm (A)}_{\sigma}(-x, x', t - t') \nonumber\\
&=& - D^{\rm (A)}_{\sigma}(x, -x', t - t')~.
\end{eqnarray}
Using the Lehmann representation one can also prove the following identity:
\begin{equation}
D^{\rm (A)}_{\sigma}(x, x', t) = D^{\rm (A)}_{\sigma}(x, x', T/2-t)~.
\end{equation}
Using these identities one can prove that the first-order term vanishes in correspondence of the extrema of $X_{{\rm CM}, \sigma}(t)$ ({\it i.e.}, for $t = n \pi/\omega$). We thus conclude that the envelope of $X_{{\rm CM}, \sigma}(t)$ decays quadratically in time, as shown in Fig.~\ref{fig:two}.

\section{The two-body problem}
\label{appendix:twobody}

The problem of two particles interacting with a short-range potential in a harmonic trap is exactly solvable~\cite{busch,polini}. This can indeed be reduced to a single-particle problem 
by switching to center-of-mass $R = (x_1+x_2)/2$ and relative motion $r = x_1-x_2$ coordinates.  The first-quantized Hamiltonian in reduced units reads ($\hbar=1$)
\begin{equation}\label{eq:ham}
{\hat {\cal H}} = -\frac{1}{4}\frac{\partial^2}{\partial R^2}+R^2 -\frac{\partial^2}{\partial r^2} + \frac{1}{4}r^2 + \tilde{g}~\delta(r)~.
\end{equation}
Distances have been rescaled with the harmonic oscillator length $a_{\rm ho} = (m\omega)^{-1/2}$ and energies with $\omega$. 
In these units $\tilde{g} = g /(\omega a_{\rm ho})$ is the dimensionless coupling constant that controls the strength of interactions.

The initial state can be factorized as
\begin{equation}\label{eq:initialstatefactorized}
\langle x_1,x_2|\psi(0)\rangle = \left(\frac{2}{\pi}\right)^{\frac{1}{4}}e^{-R^2}\times \frac{1}{(2\pi)^{\frac{1}{4}}}e^{-\frac{1}{4}(r-2d)^2}~,
\end{equation}
where $d$ is the initial displacement (as in Sec.~\ref{appendix:PT}) in units of $a_{\rm ho}$. 
The center-of-mass portion of the wave function has a trivial dynamics, while the evolution of the relative portion can be calculated numerically to the desired degree of accuracy by expanding the second factor in Eq.~(\ref{eq:initialstatefactorized}) in the exact eigenstates~\cite{busch,polini} of the Hamiltonian~\eqref{eq:ham}. 

In Fig.~\ref{fig:one-SOM} we illustrate $X_{{\rm CM}, \sigma}(t)$ as a function of time $t$ for small values of $\tilde{g}$, while in Fig.~\ref{fig:two-SOM} we show the same quantity for large values of ${\tilde g}$. These plots have to be compared with Figs.~\ref{fig:two}-\ref{fig:three}.

Comparing Fig.~\ref{fig:one-SOM}b) with Fig.~\ref{fig:two-SOM}b) we see that the time-evolution of the peak position $\left. X_{{\rm CM}, \sigma}(t)\right|_{\rm peak}$ can be fitted by a functional form which is {\it quadratic} in time at small times both in the weak- and strong-coupling regimes. Indeed, to fit the data in these panels we have used the following function, 
$\left. X_{{\rm CM}, \sigma}(t)\right|_{\rm peak} = X_0~\cos{(\sqrt{2}~t/\tau_{\rm STSD})}$, which at small times reduces to 
$\left. X_{{\rm CM}, \sigma}(t/\tau_{\rm STSD} \to 0)\right|_{\rm peak} = X_0~[1- (t/\tau_{\rm STSD})^2]$. This is at odd with the many-particle case analyzed in Figs.~\ref{fig:two}-\ref{fig:three}. 
We remind the reader that in this case the functional dependence of $\left.X_{{\rm CM}, \sigma}(t)\right|_{\rm peak}$ at strong coupling on time is linear rather than quadratic.

In Fig.~\ref{fig:three-SOM} we illustrate the dependence of $\tau_{\rm STSD}$ on the coupling constant ${\tilde g}$. In the weak-coupling regime we find $1/\tau_{\rm STSD} \propto {\tilde g}$ while in the strong-coupling regime we find $1/\tau_{\rm STSD} \propto 1/{\tilde g}$. This is in perfect agreement with the numerical results shown in Fig.~\ref{fig:four}. Note also that the order of magnitude of $\tau_{\rm STSD}/T$ in Fig.~\ref{fig:three-SOM} is the same as in Fig.~\ref{fig:four}.

\begin{figure}
\centering
\includegraphics[width=1.00\linewidth]{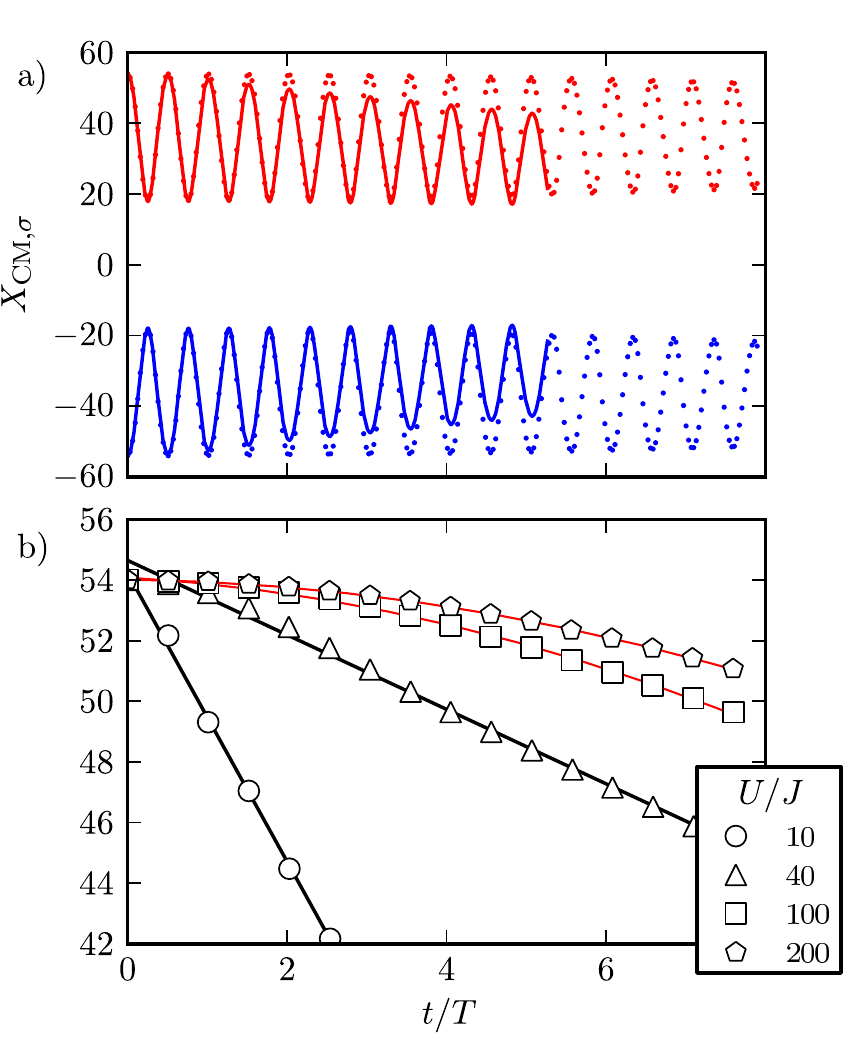}
\caption{(Color online) Same as in Figs.~\ref{fig:two}-\ref{fig:three} but for the effective ``$t$-$J$" model defined in Eq.~(\ref{eq:tJ}). 
In panel a) the solid lines correspond to $U/J =20$, while the dotted ones to $U/J = \infty$.
In panel b) the solid lines are quadratic fits of the form $\left.X_{{\rm CM}, \sigma}\right|(t)_{\rm peak} = X_0(1+bt+at^2)$. We indicated with red curves (data for $U/J=100, \, 200$) the cases in which the quadratic term in the fit is not negligible.
\label{fig:four-SOM}}
\end{figure}
\section{Split-fit procedure for the linear-to-quadratic crossover}
\label{appendix:tstar}
The mapping of the Fermi-Hubbard model into the effective ``$t$-$J$" model (Eq.~(\ref{eq:tJ})) allows to explore a wider range of values of the coupling constant $U/J$. Typical results for the spin dynamics of this model are shown in Fig.~\ref{fig:four-SOM}. Notice that for very large couplings ($U/J \approx 100-200$) the data for the peak position $\left.X_{{\rm CM}, \sigma}(t)\right|_{\rm peak}$ are not well fitted by a linear function of time, and a quadratic term becomes non-neglibible. Therefore, the peculiar linear decay of the oscillation amplitude that is visible in Fig.~\ref{fig:three}b) seems to disappear (or, better, to take place at later times) for very strong couplings.

\begin{figure*}[t]
\begin{center}
\includegraphics[scale = 1]{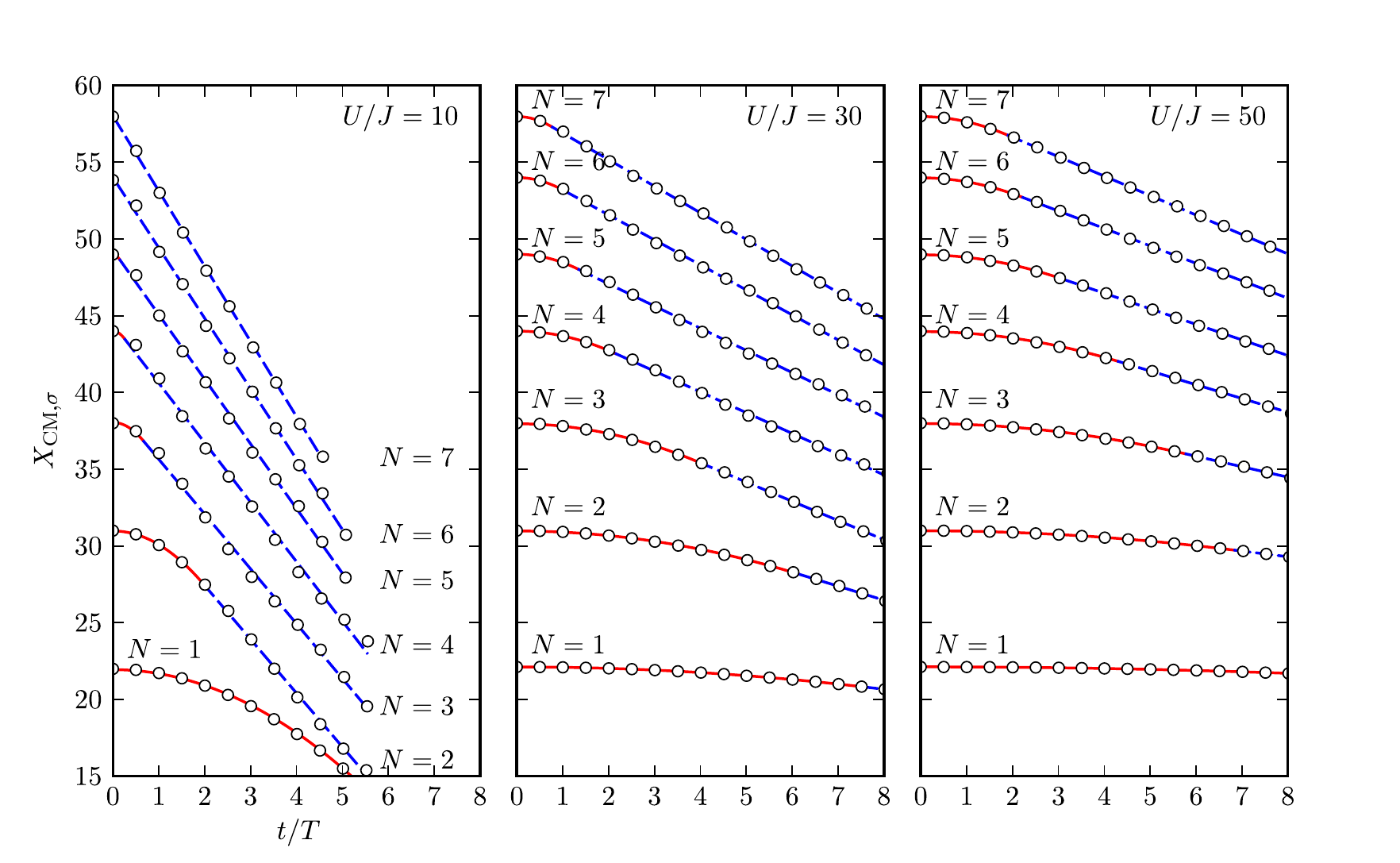}
\caption{(Color online) Time evolution of the spin-resolved center-of-mass $\left.X_{{\rm CM}, \sigma}(t)\right|_{\rm peak}$ of a system of $2N$ fermions in a harmonic potential. From top to bottom $N$ decreases from $7$ to $1$. The three different panels refer to different values of the coupling constant $U/J$. The fitting function is given in Eq.~\eqref{eq:splitfit}. The initial quadratic decay for $0 \leq t \leq t^\star$ is shown by solid red lines, while the subsequent linear decay for $t > t^\star$ by dashed blue lines. Data in this figure have been produced by employing the effective ``$t$-$J$" model - Eq.~(\ref{eq:tJ}). \label{fig:five-SOM}}
\end{center}
\end{figure*}

We have thus carried out a detailed numerical analysis of the linear-to-quadratic crossover, both as a function of the number of fermions with a given spin $N$ and of the coupling constant $U/J$, for both Fermi-Hubbard and ``$t$-$J$" models.
Our procedure is based on the key observation that the quadratic behavior takes place from $t = 0$ to a {\it crossover time} $t = t^\star(N,U/J)$, after which the oscillations decay linearly. 
This is clearly seen in Fig.~\ref{fig:five-SOM}, where we illustrate the time evolution of the spin-resolved center-of-mass $\left.X_{{\rm CM}, \sigma}(t)\right|_{\rm peak}$ for different values of $N$ and $U/J$ and for the effective ``$t$-$J$" model.

As a matter of fact, we have adopted a systematic ``split-fit" procedure by fitting our data at strong coupling with the following function:
\begin{equation}\label{eq:splitfit}
\left. X_{{\rm CM}, \sigma}(t)\right|_{\rm peak} = 
\left\{
\begin{array}{l}
{\displaystyle X_0\left(1-\frac{t^2}{2t^\star \tau_{\rm STSD}}\right)},~t \leq t^\star\vspace{0.2 cm}\\
{\displaystyle X_0\left(1-\frac{t}{\tau_{\rm STSD}}+\frac{t^\star}{2\tau_{\rm STSD}}\right)},~t > t^\star
\end{array}
\right.
\end{equation}
which yields both $\tau_{\rm STSD}$, the short-time spin-drag time constant defined in Sec.~\ref{sec:numerical_results}, and $t^\star(N,U/J)$.
The r.h.s. of Eq.~\eqref{eq:splitfit} has been expressly written in such a way to guarantee that $\left. X_{{\rm CM}, \sigma}(t)\right|_{\rm peak}$ 
is a continuous function of time $t$, with continuous first derivative, at $t= t^\star$.

\begin{figure*}[t]
\begin{center}
\includegraphics[scale = 1]{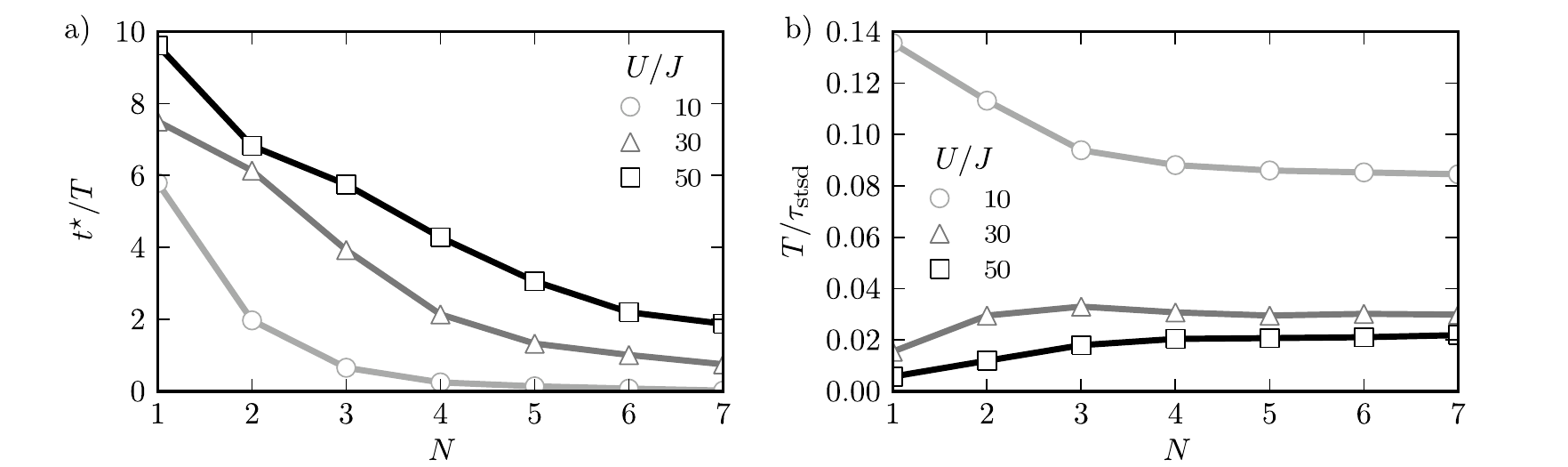}
\caption{(Color online) Dependence of $t^\star$ [panel a)] and $\tau_{\rm STSD}$ [panel b)] on $U/J$ and $N$ for the effective ``$t$-$J$" model. These data have been extracted by applying the split-fit procedure (\ref{eq:splitfit}) to the data reported in Fig.~\ref{fig:five-SOM}.\label{fig:six-SOM}}
\end{center}
\end{figure*}

From Fig.~\ref{fig:five-SOM} we clearly see that, as $N$ is increased or $U/J$ is decreased, the time scale $t^\star$, below which $\left.X_{{\rm CM}, \sigma}(t)\right|_{\rm peak}$ is well fitted by a parabolic function, decreases. 
The values of $t^\star$ and $\tau_{\rm STSD}$ extracted from this procedure are reported in Fig.~\ref{fig:six-SOM} and also in Fig.~\ref{fig:four}.


\begin{figure*}[t]
\begin{center}
\includegraphics[scale = 1]{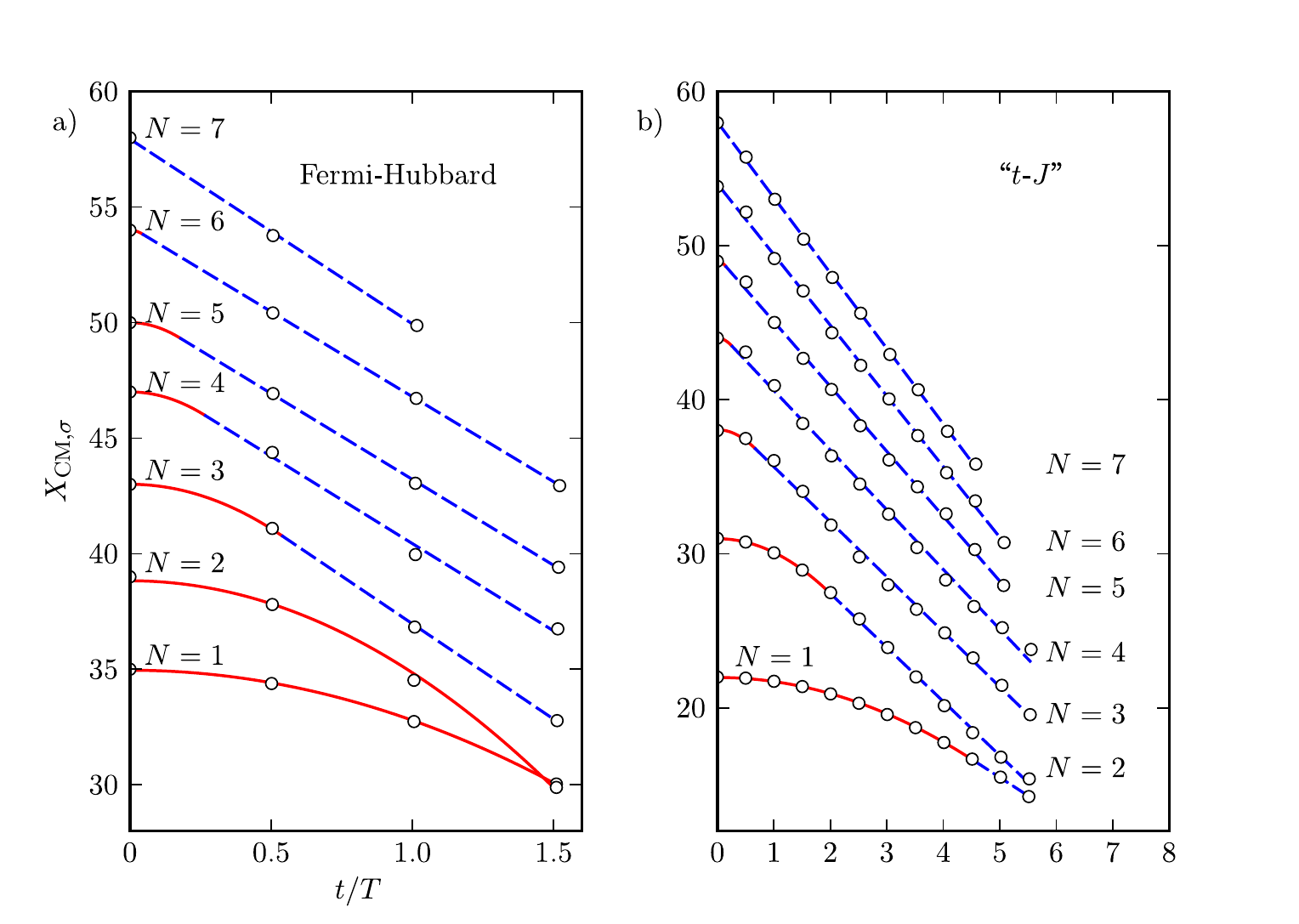}
\caption{(Color online) Same as in Fig.~\ref{fig:five-SOM}, but only for $U/J =10$. Panel a) refers to the Fermi-Hubbard model, while panel b) refers to the effective ``$t$-$J$" model. Note the different horizontal scale: the simulation of long-time spin dynamics of the Fermi-Hubbard model is more difficult.\label{fig:seven-SOM}}
\end{center}
\end{figure*}

\begin{figure*}[t]
\begin{center}
\includegraphics[scale = 1]{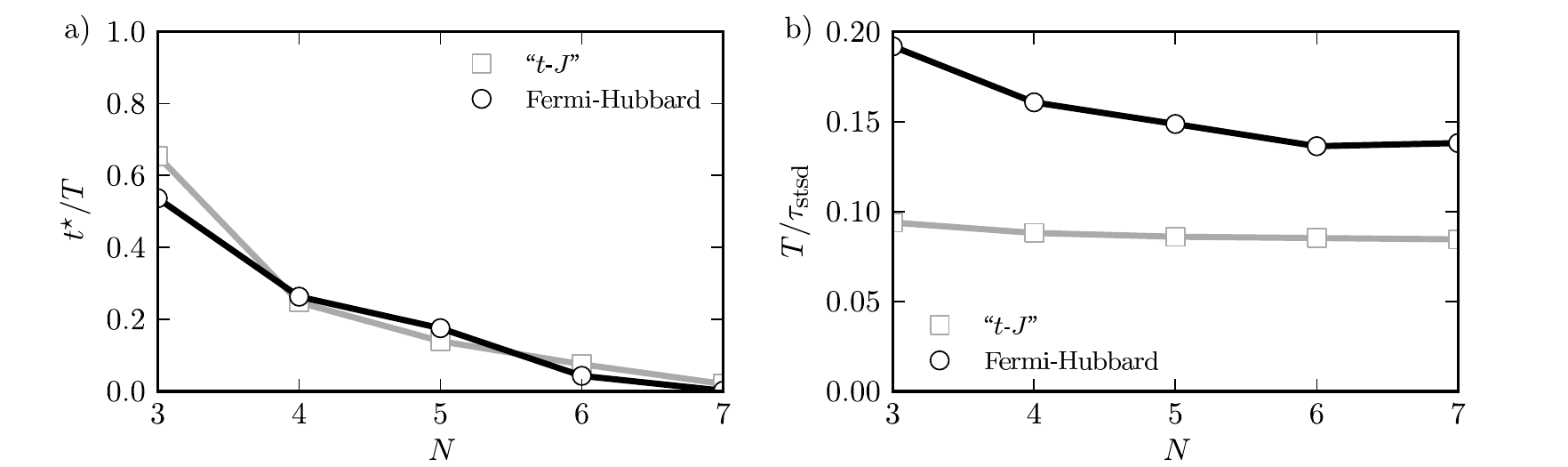}
\caption{(Color online) Same as in Fig.~\ref{fig:six-SOM}, but only for $U/J=10$. Here we compare results 
for the Fermi-Hubbard model [Fig.~\ref{fig:seven-SOM}a)] with those for the effective ``$t$-$J$" model [Fig.~\ref{fig:seven-SOM}b)]. Note that the values of $t^\star$ for the two models are practically identical.\label{fig:eight-SOM}}
\end{center}
\end{figure*}

We have repeated the same numerical analysis also for the Fermi-Hubbard model. The results are reported in Figs.~\ref{fig:seven-SOM}-\ref{fig:eight-SOM} and also in Figs.~\ref{fig:three}-\ref{fig:four}.



\begin{thebibliography}{99}
\bibitem{foot1}
Here $\langle {\hat n}_{i, \sigma}(t)\rangle$ is a shorthand for $\langle\Psi(t)|{\hat n}_{i, \sigma}|\Psi(t)\rangle$, where $|\Psi(t)\rangle$ is the state of the system at time $t$.
\bibitem{foot2}
To achieve the continuum limit we need to consider a rather large number of sites and a sufficiently shallow harmonic confinement.
        This forces the quantum dynamics to be slow on the time scale of $\hbar / J$. 
        As a matter of fact, with the parameters used throughout this Letter, the period induced 
        by the harmonic confinement is $T = 160 \pi (\hbar / J)$.
        Despite the intrinsic difficulty of tDMRG in addressing the long-time $t \gg \hbar/J$ limit, 
        we are able to follow the dynamics up to $t \sim 6~T$ 
        thanks to the small number of particles ($2N \ll L$).
\bibitem{mapping}
	The mapping ${\hat {\cal H}} \to {\hat {\cal H}}'$ for $U/J \gg 1$ is, however, not exact since in writing Eq.~(\ref{eq:tJ}) we have neglected 
	three-site terms, which are difficult to simulate and potentially important away from half filling~\cite{essler_book}.
\end{thebibliography}
\end{document}